\documentclass{epl}
\usepackage{amssymb}
\usepackage{amsmath}
\title{Breakdown of the Sonine expansion \\ 
  for the velocity distribution of Granular Gases}
\shorttitle{Breakdown of the Sonine expansion\dots}

\author{Nikolai V. Brilliantov\inst{1,2} \and Thorsten P\"oschel\inst{3}}
\shortauthor{N.V. Brilliantov \and T. P\"oschel}
\institute{
  \inst{1}Institute of Physics, University of Potsdam, 
  Am Neuen Palais 10, 14469 Potsdam, Germany \\
  \inst{2} Department of Physics, Moscow State University,  
  Vorobievy Gory 1, 119899 Moscow, Russia \\
  \inst{3}Institut f\"ur Biochemie, Charit\'e, 
  Monbijoustra{\ss}e 2, 10117 Berlin, Germany 
}


\begin{document}

\maketitle

\begin{abstract}
  The velocity distribution of a granular gas is analyzed in terms of
  the Sonine polynomials expansion. We derive an analytical expression
  for the third Sonine coefficient $a_3$. In contrast to frequently
  used assumptions this coefficient is of the same order of magnitude
  as the second Sonine coefficient $a_2$. For small inelasticity the
  theoretical result is in good agreement with numerical simulations.
  The next-order Sonine coefficients $a_4$, $a_5$ and $a_6$ are
  determined numerically. While these coefficients are negligible for
  small dissipation, their magnitude grows rapidly with increasing
  inelasticity for $0< \varepsilon \lesssim 0.6$. We conclude that
  this behavior of the Sonine coefficients manifests the break down of
  the Sonine polynomial expansion caused by the increasing impact of
  the overpopulated high-energy tail of the distribution function.
\end{abstract}

\section{Introduction}

The velocity distribution function of granular gases deviates from the
Maxwell distribution, as first described by Goldshtein and Shapiro
\cite{GoldshteinShapiro1:1995}. This deviation depends on the
coefficient of restitution $\varepsilon$, which quantifies the loss of
energy for a collision of two particles $i$ and $j$:
\begin{equation}
  \label{eq:eps}
  \vec{v}_i^{\,\prime} = \vec{v}_i 
  - \frac{1+\varepsilon}{2}\left[\left(\vec{v}_i-\vec{v}_j\right)\cdot \vec{e} \, 
  \right]\vec{e}\,,~~~~~~
  \vec{v}_j^{\,\prime} = \vec{v}_j + \frac{1+\varepsilon}{2}\left[\left(\vec
      {v}_i-\vec{v}_j\right)\cdot
    \vec{e}\, \right]\vec{e}\,.
\end{equation}
Here $\vec{v}_i^{\,\prime}$ and $\vec{v}_j^{\,\prime}$ stand for the
post-collisional velocities where the unit vector of the relative
particle position at the collision instant is $\vec{e}
\equiv\left(\vec{r}_i-\vec{r}_j\right)/\left|\vec{r}_i-\vec{r}_j\right|$.

The deviation from the Maxwell distribution may be described by a
Sonine polynomials expansion
\cite{GoldshteinShapiro1:1995,NoijeErnst:1998,BrilliantovPoeschelStability:2000}.
This expansion is applicable to the main part of the velocity
distribution, excluding the high-energy tails, which is known to be
exponentially overpopulated \cite{EsipovPoeschel:1995}.

So far it was silently accepted that the Sonine expansion is a
converging series while the exponential tail does not noticeably
contribute to the coefficients of this expansion. This assumption was
supported by Direct Simulation Monte Carlo (DSMC) of the Boltzmann
equation \cite{BreyCuberoRuizMontero:1996}, as well as by Molecular
Dynamics simulations of Granular Gases \cite{HuthmannOrzaBrito:2000}.
Up to now, however, neither the region of validity of the Sonine
expansion is known, nor the impact of the exponential tail on the
convergence of this series.

Huthmann, Orza, and Brito \cite{HuthmannOrzaBrito:2000} derived a
system of equations for the Sonine coefficients from the Boltzmann
equation and solved it perturbatively with the assumption that the
Sonine coefficient $a_k$ is of the order ${\cal O}
\left(\lambda^k\right)$ with $\lambda$ being a small parameter. For
$\varepsilon \gtrsim 0.6$ they obtained rapid decrease of the Sonine
coefficients with escalating order $k$, while for smaller values of
$\varepsilon$ the high-order coefficients, $k \ge 3$, were not
negligible and could be of the same order as the first non-trivial
coefficient $a_2$. It was not evident, however, whether the Sonine
polynomials expansion breaks down for $\varepsilon \lesssim 0.6$, or
the perturbative approach based on the conjecture $a_k \sim \lambda^k$
was inadequate.

In the present study we derive the third Sonine coefficient, $a_3$,
and address the convergence of the Sonine polynomials expansion
analytically and numerically by means of DSMC. We show that the Sonine
coefficients do not decrease with increasing $k \ge 3$ for
$\varepsilon \lesssim 0.6$, i.e., the Sonine series diverges. We
conclude that the breakdown of the Sonine expansion is caused by the
increasing impact of the exponentially overpopulated tail for large
dissipation.

\section{Sonine polynomials expansion}

We consider a Granular Gas of particles of mass $m$ which interact
with a constant coefficient of restitution $\varepsilon$=const. and
neglect their rotational degrees of freedom. We assume that the gas is
in the homogeneous cooling state at number density $n=N/V$. Moreover,
we assume that the velocity distribution function of the gas
$f\left(\vec{v}, t\right)$ has acquired its scaling form
\cite{EsipovPoeschel:1995,GoldshteinShapiro1:1995}, i.e.
\begin{equation}
\label{eq:defScalf}
f\left(\vec{v},t\right)=\frac{n}{v_T^d(t)} \tilde{f}\left(\vec{c}\,\right)\,,
\qquad\qquad \vec{c} \equiv \frac{\vec{v}}{v_T(t)} \,, 
\end{equation}
where\,$d$\,is the system dimension and $v_T(t)$ is the thermal
velocity due to the temperature $T(t)$,
\begin{equation}
  \label{eq:Tdef}
  \frac{d}{2}n T =   \int \frac{mv^2}{2}f(v) d \vec{v}  
                 = \frac{d}{2}n \, \frac{mv_T^2}{2} \,. 
\end{equation}
For $\varepsilon ={\rm const.}$ the Boltzmann equation reduces to two
uncoupled equations, for the temperature and for the reduced
distribution function $\tilde{f}(c)$
\cite{GoldshteinShapiro1:1995,EsipovPoeschel:1995}. The first equation
reads
\begin{equation}
\label{eq:dTdt}
\frac{dT}{dt}= -\frac{2}{d}g_2(\sigma) \sigma^{d-1} n v_T T \mu_2 \, , 
\end{equation}
where $\sigma$ is the particle diameter. For $d=3$, the contact value
of the pair distribution function is given by $g_2(\sigma) = \left(1-
  \eta/2\right)/(1-\eta)^3$ with the packing fraction $\eta= n \pi
\sigma^3/6$, and
\begin{equation}
\label{eq:def:mu2}
\mu_p \equiv - \int d \vec{c}_1 c_1^{\,p} \tilde{I}(\tilde{f},\tilde{f}) \,  
\end{equation} 
denotes the moments of the dimensionless collision integral
\cite{NoijeErnst:1998,BrilliantovPoeschelStability:2000,BrilliantovPoeschel:2000visc},
\begin{equation}
  \label{eq:def:dimlI}
  \tilde{I}\left(\tilde{f},\tilde{f}\right)  \equiv \int d \vec{c}_2 \int d 
  \vec{e}\, \Theta\left(-\vec{c}_{12} \cdot \vec{e}\,\right) 
  \left|\vec{c}_{12} \cdot \vec{e}\,\right| \, 
  \,  \left[ \frac{1}{\varepsilon^2} \tilde{f}(\vec{c}_1^{\, \prime \prime}) 
    \tilde{f}(\vec{c}_2^{\, \prime \prime}) - \tilde{f}(\vec{c}_1)
    \tilde{f}(\vec{c}_2) \right] \, . 
\end{equation}
The unit step function $\Theta(x)$ guarantees that only approaching
particles collide, $\left|\vec{c}_{12} \cdot \vec{e}\,\right| $ gives
the length of the collision cylinder and $\vec{c}_1^{\,\prime \prime}$
and $\vec{c}_2^{\, \prime \prime}$ denote the reduced velocities for
the inverse collision, i.e., for the collision which results at the
reduced velocities $\vec{c}_1$ and $\vec{c}_2$.

The second equation for the reduced distribution function reads 
\begin{equation}
  \label{eq:fviaI}
  \frac{\mu_2}{d}\left( d + c_1 \frac{\partial}{\partial c_1} \right) 
  \tilde{f}\left(\vec{c}_1\right) =
\tilde{I}(\tilde{f},\tilde{f}) \,.  
\end{equation} 

For the case of elastic collisions, $\varepsilon=1$, the resulting
velocity distribution is the Maxwell distribution. Therefore, for
sufficiently large coefficient of restitution, $\varepsilon \to 1$, we
may assume that $\tilde{f}(c)$ is close to the Maxwellian,
$\phi(c)=\pi^{d/2} \exp \left(-c^2\right)$. This suggests to expand
the (unknown) distribution function $\tilde{f}(c)$ around $\phi(c)$ in
terms of orthogonal polynomials $S_p(x)$:
\begin{equation}
  \label{eq:Sonexp}
  \tilde{f}(c)=\phi(c) \varphi(c) = \phi(c)\left[ 1 
    + \sum_{p=1}^{\infty} a_p S_p\left(c^2\right) \right]  \,.
\end{equation} 
We chose $S_p(x)$ to be the Sonine polynomials (see, e.g., 
\cite{BrilliantovPoeschelOUP}):
\begin{equation}
\label{eq:Sondef}
S_p(x)= \sum_{n=0}^{p} \frac{(-1)^n (p+1/2)!}{(n+1/2)!(p-n)!n!}x^n  \,.  
\end{equation} 
The first few of them, relevant for this study,  read 
\begin{equation}
\label{eq:Sonindef}
  \begin{split}
    S_1(x) & = -x+\frac12 d\\
    S_2(x) & =\frac12 x^2 - \frac12(d+2)x +\frac18 d(d+2)\\
    S_3(x) & =-\frac16 x^3 +\frac14 (d+4)x^2 -\frac18 (d+2)(d+4)x 
    +\frac{1}{48} d(d+2)(d+4)\,.     
  \end{split}
\end{equation}
The expansion coefficients $a_k$ characterize the deviation of the
distribution function from the Maxwell distribution, namely, they
quantify the deviation of the moments of the distribution function,
$\left< c^p \right> \equiv \int d\vec{c} c^p \tilde{f}(c)$, from the
corresponding values for the Maxwell distribution. The equations for
the moments may be found multiplying both sides of Eq.
\eqref{eq:fviaI} by $c_1^p$, integrating over $ \vec{c}_1$, and using
the orthogonality of the Sonine polynomials. This yields an infinite
set of equations
\cite{NoijeErnst:1998,BrilliantovPoeschelStability:2000,BrilliantovPoeschelOUP},
\begin{equation}
\label{eq:mu2mup}
d \, \mu_p = \mu_2 \, p \, \left< c^p \right>  \,,
\qquad p=2,4, \ldots   
\end{equation} 
Since $\left< c^p \right>$ and $\mu_p$ are expressed in terms of the
Sonine coefficients, this set of equations can be used to determine
the Sonine coefficients and, thus, to find the velocity distribution.
To close the set of equations, a cutoff of the series
\eqref{eq:Sonexp} is applied, that is, it is assumed that the Sonine
coefficients $a_k$ with $k>k_0$ are negligible.

The first Sonine coefficient vanishes, $a_1=0$, according to the
definition of temperature \cite{GoldshteinShapiro1:1995}. Since
$\left< c^2 \right> =\frac12 d$, the first equation in Eq.
\eqref{eq:mu2mup} for $p=2$ gives the identity. The second equation in
Eq. \eqref{eq:mu2mup} for $p=4$ allows to find $a_2$ by expressing
$\left< c^4 \right>$, $\mu_2$ and $\mu_4$ in terms of $a_2$ and
neglecting all other Sonine coefficients $a_3$, $a_4, \ldots$
\cite{GoldshteinShapiro1:1995}. Therefore, the first non-trivial
Sonine coefficient is $a_2$.

\section{Second and third Sonine coefficients}

We write $\tilde{f}(c)=\phi(c) \left[1 + a_2 S_2(c^2) + a_3
  S_3(c^2)\right]$ for the velocity distribution function by assuming
that $a_4$, $a_5, \ldots$ are negligible. Further, we write Eq.
\eqref{eq:mu2mup} for $p=4$ and $p=6$ and express all quantities in
these two equations in terms of $a_2$ and $a_3$. In particular,
\begin{equation}
  \label{eq:c4c6}
  \left< c^4 \right> = \frac14 d (d+2)(1+a_2) \, ,  
  \qquad
  \left< c^6 \right> = \frac18 d (d+2)(d+4) (1+3a_2-a_3) \,.   
\end{equation} 
To find the moments $\mu_2$, $\mu_4$, and $\mu_6$ we recast Eq.
\eqref{eq:def:mu2} using the properties of the collision integral
(see, e.g., \cite{BrilliantovPoeschelOUP}):
\begin{equation}
  \label{eq:mupviaDelta}
  \mu_p = -\frac12 \int d \vec{c}_1 \int d \vec{c}_2 \int d \vec{e}\, 
  \Theta\left(-\vec{c}_{12} \cdot \vec{e}\,\right) \left|\vec{c}_{12} 
    \cdot \vec{e}\, \right| \tilde{f}(c_1)\tilde{f}\left(c_2\right) 
  \Delta\left(c_1^p+c_2^p\right) \,, 
\end{equation}
where $\Delta \psi(\vec{c}_i)= \psi\left(\vec{c}_i^{\,
    \prime}\right)-\psi\left(\vec{c}_i\,\right)$ denotes the variation
of some quantity $\psi\left(\vec{c}\right)$ in a direct collision. The
evaluation of integrals of the form of Eq. \eqref{eq:mupviaDelta} has
been described in detail in \cite{BrilliantovPoeschelOUP}. Using this
approach, analytical expressions for $\mu_2$, $\mu_4$, and $\mu_6$ may
be obtained, which are rather cumbersome.  Since it is expected that
$a_2 \gg a_2^2$, $a_3\gg a_3^2$, $a_2\gg a_2a_3$, and $a_3\gg a_2a_3$,
we keep in these moments only linear terms with respect to $a_2$ and
$a_3$:
\begin{equation}
  \label{eq:mu4lin}
  \begin{split}
    \mu_2 &= \frac{\pi^{d/2}}{\sqrt{2 \pi} \Gamma(d/2)} 
       \left( 1 -\varepsilon^2 \right) \left[1 + \frac{3}{16}a_2 + \frac{1}{64} a_3 \right]\\
    \mu_4 &= \frac{\pi^{d/2}}{\sqrt{2 \pi} \Gamma(d/2)} 
       \left[T_1 + T_2a_2 + T_3a_3 \right] \\
    \mu_6 &= \frac{\pi^{d/2}}{\sqrt{2 \pi} \Gamma(d/2)} 
       \left[D_1 + D_2a_2 + D_3a_3 \right] 
  \end{split}
\end{equation}
where
\begin{equation}
  \label{eq:T2}
  \begin{split}
    T_1 &= \left( 1 -\varepsilon^2 \right)  
       \left( d + \frac32 + \varepsilon^2 \right) \\ 
    T_2 &= \frac{3}{32} \left( 1 -\varepsilon^2 \right) 
       \left( 10d +10\varepsilon^2 +39\right)+\left( 1 + \varepsilon \right) 
       \left(d -1 \right) \\
    T_3 &= \frac{(1- \varepsilon^2)}{128} 
       \left( 10 \varepsilon^2 +97 \right)
       -\frac{\left( 1 + \varepsilon \right) \left(d -1 \right)}{64} 
       \left( 21- 5 \varepsilon \right)\\
    D_1 &= \frac34 \left( 1 -\varepsilon^2 \right) 
       \left[ \left( d + \varepsilon^2 \right) 
         \left( 5 + 2 \varepsilon^2 \right) +  d^2+\frac{19}{4} \right] \\ 
    D_2 &= \frac{3}{256}\left( 1 -\varepsilon^2 \right) 
       \left[ 1289 - 4\left( d + \varepsilon^2 \right)
         \left( 311 + 70 \varepsilon^2 \right) +  172d^2 \right] 
       + \frac34 B(\varepsilon) \\
    D_3 &= -\frac{3}{1024}\left( 1 -\varepsilon^2 \right) 
       \left[ 2537 + 4\left( d + \varepsilon^2 \right)
         \left( 583 + 70 \varepsilon^2 \right) +  236 d^2 \right] 
       - \frac{9}{16} B(\varepsilon) \\
    B &= \left( 1 + \varepsilon \right) 
       \left[\left(d-3 \right) \left(3 + 4 \varepsilon^2 \right) + 2 
         \left( d^2 - \varepsilon \right) \right] \, .
  \end{split}
\end{equation}
For $a_3=0$, the above equations for $\mu_2$ and $\mu_4$ coincide with
the equivalent equations in
\cite{NoijeErnst:1998,BrilliantovPoeschelStability:2000}.
 
Substituting Eqs. \eqref{eq:c4c6}, \eqref{eq:mu4lin} and \eqref{eq:T2}
into Eqs. \eqref{eq:mu2mup} for $p=4$ and $p=6$ we obtain the Sonine
coefficients $a_2$ and $a_3$ in linear approximation. For $d=3$, the
result reads
\begin{equation}
  \label{eq:a3lin}
  \begin{split}
    a_2 &= -\frac{16}{b(\varepsilon)} 
       \left(240\varepsilon^8\!-480\varepsilon^7\!+3312\varepsilon^6\!-7424\varepsilon^5\!
         +3510\varepsilon^4\!-364\varepsilon^3
         +895\varepsilon^2\!+1934\varepsilon-1623\right)\\
    a_3 &= -\frac{128}{b(\varepsilon)} 
       \left(80\varepsilon^8-160\varepsilon^7+816\varepsilon^6-1600\varepsilon^5
         +154\varepsilon^4+1548\varepsilon^3
      -669\varepsilon^2-386\varepsilon+217\right)
  \end{split}
\end{equation}
\begin{multline}
  \label{eq:beps}
  b(\varepsilon)=2800\varepsilon^8-5600\varepsilon^7+34800\varepsilon^6
  -84480\varepsilon^5-4410\varepsilon^4+25716\varepsilon^3
  +112155\varepsilon^2-172458\varepsilon+214357. 
  \nonumber
\end{multline}

\section{DSMC simulations}

To check the predictions of the theory and to study the behavior of
higher Sonine coefficients we perform Monte Carlo simulations of the
Boltzmann equation (DSMC) using $2\times 10^7$ particles of unit mass.
The coefficient of restitution was varied in the interval
$\varepsilon\in (0.1, 1)$ in steps of $0.01$.  To obtain smooth data,
for each value of $\varepsilon$ we performed 80 simulations and
recorded the velocities of the particles when the system had reached a
state with a scaling distribution function, Eq. \eqref{eq:defScalf}.
From these snapshots we computed the temperature and the moments of
the reduced distribution functions evaluating the averages
\begin{equation}
  \label{eq:ckT}
  T= \frac{1}{3N}\sum_{i=1}^N\vec{v}_i \cdot\vec{v}_i \, , 
  \qquad \qquad 
  \left<c^{2k}\right>= \frac{1}{N} \, \sum_{i=1}^N\left(\frac{\vec{v}_i \cdot\vec{v}_i}{2T}\right)^k  \,.
\end{equation}
The Sonine coefficients can be computed using the relation (see e.g.
\cite{BrilliantovPoeschelOUP})
\begin{equation}
  \label{eq:ckak}
  \left<c^{2k}\right>=\frac{(2k+1)!!}{2^k} 
  \left( 1+ \sum_{p=1}^{k} (-1)^p \frac{k!}{(k-p)! p! } a_p \right) \, .
\end{equation}

\begin{figure}[t!]
  \centerline{\includegraphics[width=8cm,clip]{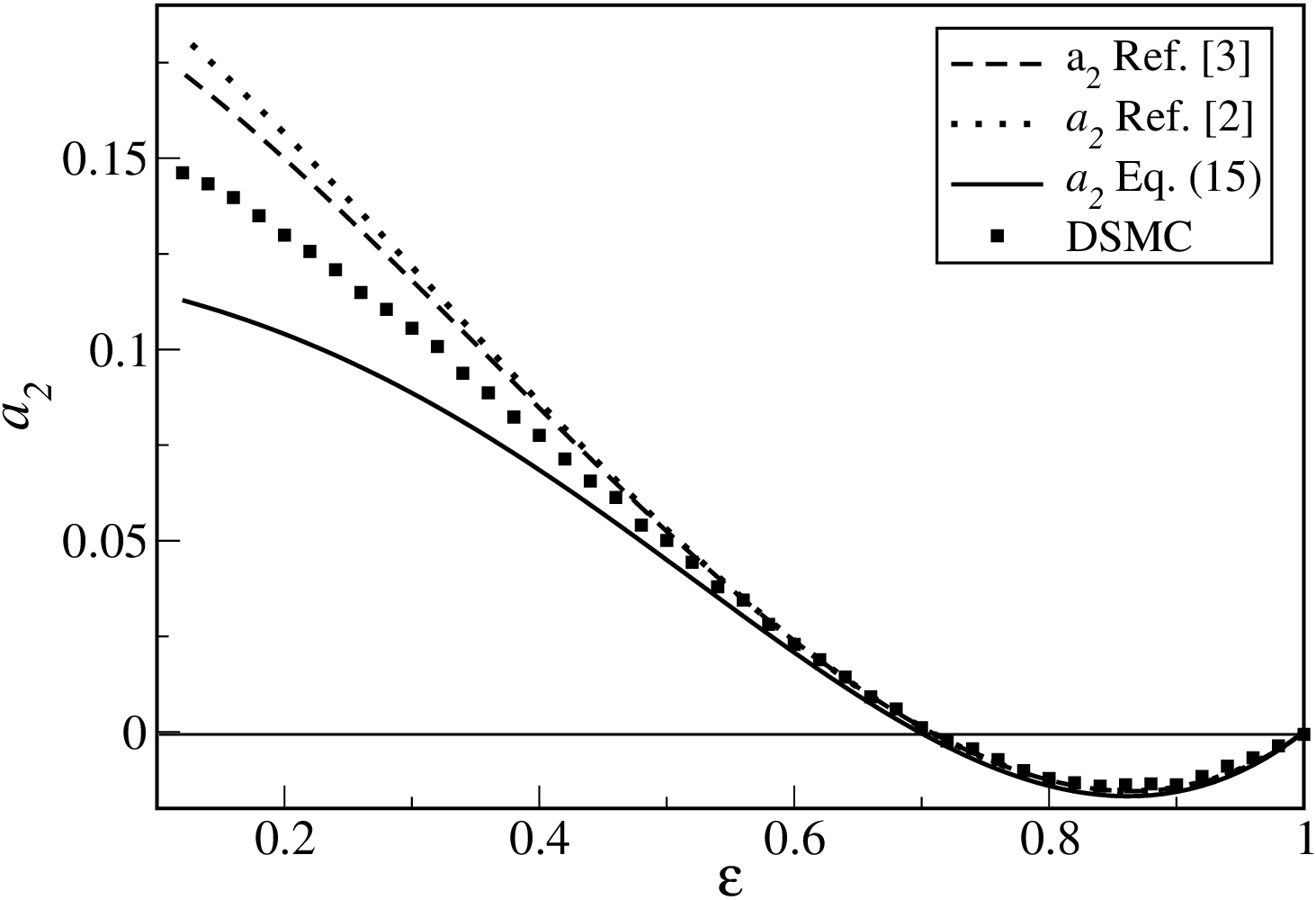}}
  \caption{The second Sonine coefficient $a_2$ over the coefficient of
    restitution $\varepsilon$ as given by Eq. \eqref{eq:a3lin} where
    $a_2$ and $a_3$ are taken into account, $a_2(\varepsilon)$ from
    the linear theory, where only $a_2$ is taken into account
    \cite{NoijeErnst:1998}, and $a_2(\varepsilon)$ from the
    corresponding non-linear theory
    \cite{BrilliantovPoeschelStability:2000}, together with DSMC
    results.}
\label{fig:a2}
\end{figure}

\begin{figure}[t!]
\centerline{\includegraphics[width=8cm,clip]{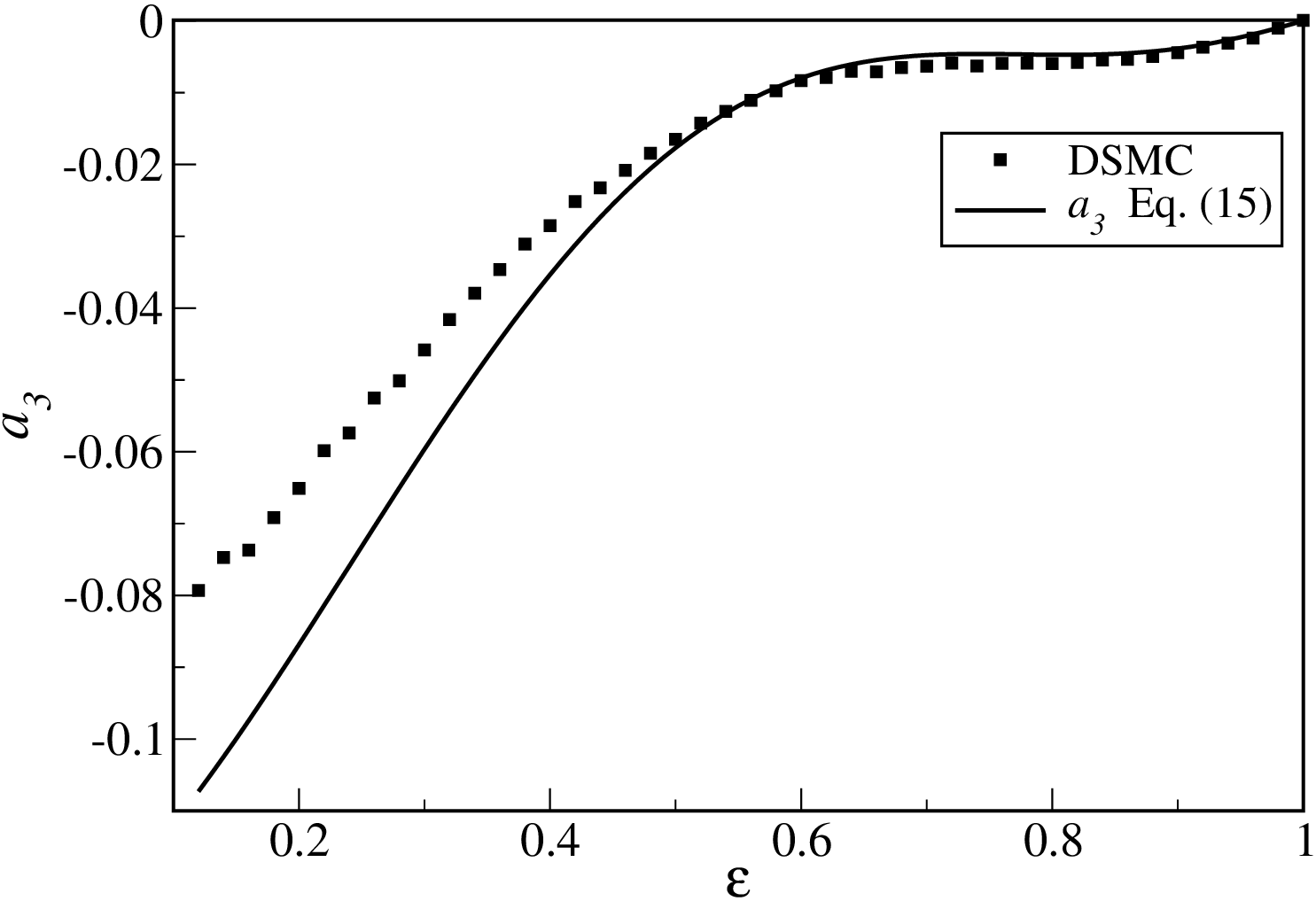}}
\caption{The coefficient $a_3$ over the coefficient of restitution
  $\varepsilon$ due to Eq. \eqref{eq:a3lin} and to DSMC.}
  \label{fig:a3}
\end{figure}

Figure \ref{fig:a2} shows $a_2$ as given by Eq. \eqref{eq:a3lin}
(which takes $a_3$ into account), $a_2$ as it follows from the linear
theory \cite{NoijeErnst:1998} and $a_2$ due to a non-linear theory
\cite{BrilliantovPoeschelStability:2000} together with DSMC results.
All approaches agree fairly well with the simulation results for small
inelasticity, $\varepsilon \lesssim 0.6$, and deviate noticeably for
larger dissipation.
Figure \ref{fig:a3} shows $a_3$ due to Eq. \eqref{eq:a3lin} together
with the DSMC results. Again we see that the predictions of the new
theory are in good agreement with the numerical results for
$\varepsilon \gtrsim0.6$.
Similar numerical investigations of $a_3$ have been done by Brey et
al. \cite{BreyCuberoRuizMontero:1996} for $\varepsilon>0.7$ and of
$a_3$, $a_4$, $a_5$ by Nakanishi \cite{Nakanishi:2003} for
$\varepsilon>0.9$.
 
\section{Higher Sonine coefficients}

\begin{figure}[t!]
  \centerline{\includegraphics[width=8cm,clip=]{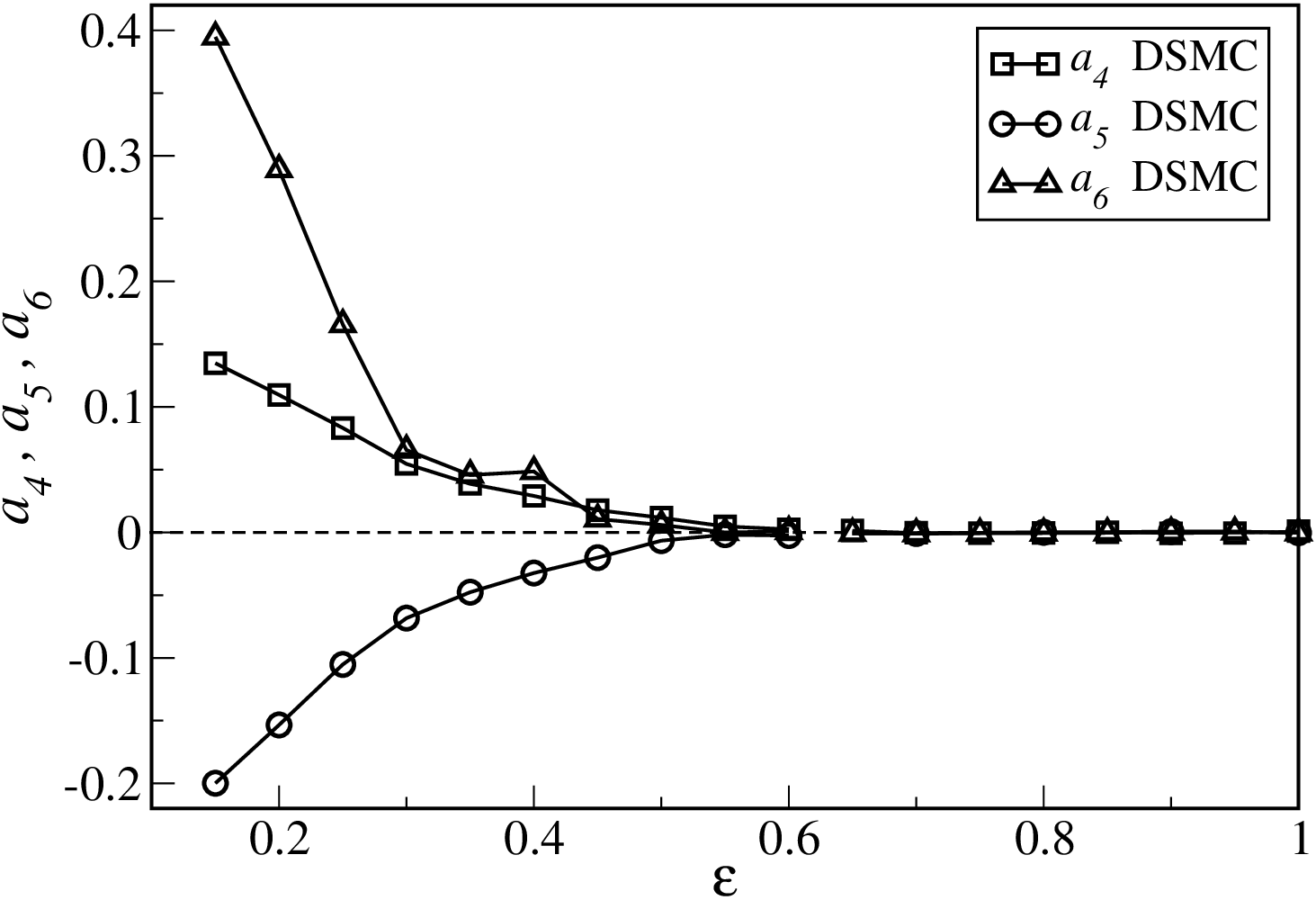}}
  \caption{High-order Sonine coefficients as functions of the
    coefficient of restitution (symbols). The lines guide the eye.}
  \label{fig:a456}
\end{figure}

In Fig. \ref{fig:a456} we present the high-order Sonine coefficients
as functions of the coefficient of restitution $a_4(\varepsilon)$,
$a_5(\varepsilon)$, and $a_6(\varepsilon)$.
These coefficients are very small for $\varepsilon \lesssim 0.6$,
which indicates the convergence of the Sonine expansion. For larger
inelasticity, however, the absolute values of $a_p$, $p=4,\, 5,\, 6$
increase with increasing $p$. Hence we conclude that for large
inelasticity, $\varepsilon\lesssim 0.6$, the Sonine expansion does not
converge.  A similar result was reported in
\cite{HuthmannOrzaBrito:2000}, where the Sonine coefficients were
found from the Boltzmann equation under the assumption $a_k \sim
\lambda^k$ ($\lambda \ll 1$).

What is the reason for the breakdown of the Sonine expansion with
increasing inelasticity? This happens due to the increasing impact of
the overpopulated tail of the velocity distribution function
\cite{EsipovPoeschel:1995}, which reads for $c \gg 1$
\cite{EsipovPoeschel:1995,NoijeErnst:1998}:
\begin{equation}
  \label{eq:tail}
  \tilde{f}(c) \sim  B e^{-bc} \, ; \qquad \qquad 
  b= \frac{\pi^{(d-1)/2}d}{\Gamma\left(\frac{d+1}{2} \right) \mu_2}  \, , 
\end{equation}
where $\mu_2$ has been defined above, while the prefactor $B$ is
unknown. For small $\varepsilon$ the exponential overpopulation starts
for velocities that are not significantly larger than the thermal
velocity. In this case the contribution to the moments $\left< c^{2k}
\right>$ from the exponential tail rapidly grows with increasing $k$,
which entails the corresponding growth of $a_k$ and ultimately, the
breakdown of the Sonine expansion. Simulations show that at least in
some range of $\varepsilon$ the main part of the distribution with $c
\sim 1$ has a crossover to the tail distribution, Eq. \eqref{eq:tail},
at $c \sim c^*$ \cite{tail}. Assuming that the overpopulation of the
tail starts at $c^* \sim b \sim 1/(1- \varepsilon^2)$
\cite{BrilliantovPoeschelOUP} one can analyze the contribution from
the exponential tail to the moments $\left< c^{2k} \right>$.

The breakdown of the Sonine expansion may be also understood from a
simple mathematical argument: The Sonine expansion, Eq.
\eqref{eq:Sonexp}, contains only even powers of the scaled velocity
$c$. The tail of the distribution decaying as $\exp(-c)$, $c\in
(c^*,\infty)$, however, cannot be represented by a series in even
powers of $c$. Therefore, for any value of $\varepsilon < 1$, the
presence of the asymptotic exponential tail {\em must} invalidate the
Sonine expansion in the limit $c\to\infty$.

For $\varepsilon\lesssim 1$ the tail starts at rather large velocity $c^* \gg
1$, thus, only high-order terms of the Sonine expansion which are sensitive to
large values of $c$ are affected and the corresponding high-order Sonine
coefficients $a_p$ are large. Indeed, for $c\gg 1$, according to Eq.~\eqref{eq:Sonexp}
\begin{equation}
  \label{eq:scal}
  \varphi (c) =\frac{\tilde{f}(c)}{\phi(c)} 
  = \sum_p a_pS_p(c^2) \to B \exp\left(-bc +c^2 \right) 
  \simeq B \exp\left( c^2\right) = B\sum_p \frac{c^{2p}}{p!} \,, 
\end{equation}
while the Sonine polynomials may be approximated by their
leading-order terms:
\begin{equation}
  \label{eq:sonlead}
  S_p \approx \frac{(-1)^p c^{2p}}{p!}\,,~~~~\mbox{thus}~~~~a_p\to (-1)^pB~~~ \mbox{for~~} p\gg1\,.
\end{equation}
The distribution function itself, however, is extremely small for
these velocities.  For larger inelasticity, that is, smaller
$\varepsilon$, the value of $c^*$ is not significantly larger than the
thermal velocity. In this case, as shown in the present paper, already
smaller Sonine coefficients such as $a_3$ to $a_6$ may diverge.

Nevertheless, the Sonine expansion remains a valuable tool for
describing the main part of the velocity distribution, $\tilde{f}(c)$,
for $c< c^*$ which is the range of interest in most cases. When
considering the expansion up to $a_6$, the range $\varepsilon \in
(0.6,1)$ seems to be a safe interval for applying the Sonine
expansion. For higher-order expansions, however, the range of validity
may be restricted to a smaller interval. A more quantitative
discussion will be given in \cite{tail}
 
\section{Conclusion}

We derived analytical expressions for the first two non-trivial Sonine
coefficients as functions of the coefficient of restitution,
$a_2(\varepsilon)$ and $a_3(\varepsilon)$, and show that the
coefficient $a_3$ is not negligible as compared to $a_2$ as it was
assumed in previous theories. We show that for small inelasticity,
$0.6 \lesssim \varepsilon <1$, $a_k$ with $k \ge 4$ are significantly
smaller than $a_2$, $a_3$ and may be neglected, that is, the Sonine
expansion converges. For this interval of $\varepsilon$ the obtained
theoretical values of $a_2$ and $a_3$ are in a good agreement with
numerical results. We also find numerically the coefficients $a_4$,
$a_5$, and $a_6$ for a wide range of the restitution coefficient, $
0.1\le \varepsilon \le 1$. For large inelasticity, $ 0< \varepsilon
\lesssim 0.6$, the high-order Sonine coefficient are of the same order
as $a_2$, $a_3$ and, hence, may not be neglected. The reason for this
behavior is the significant contribution of the exponentially
overpopulated tail to the moments of the velocity distribution
function. This contribution rapidly growths with increasing
inelasticity and the order of the moments, ultimately undermining the
Sonine polynomial expansion.


\end{document}